\documentstyle[aps,prl]{revtex}

\begin{document}

\twocolumn[{
\draft
\widetext

\title{Mesoscopic Capacitors: A Statistical Analysis.}
\author{V\'\i ctor A. Gopar$^1$, Pier A. Mello$^1$ and Markus B\"uttiker$^2$}

\address{$^1$Instituto de F\'\i sica, Universidad Nacional Aut\'{o}noma de
M\'exico\\
01000 M\'exico, D.F., M\'exico}
\address{$^2$D\'epartement de Physique Th\'eorique, Universit\'e de Gen\`eve,
CH-1211 Gen\`eve 4, Switzerland}

\date{Submitted to Phys Rev. Lett., April 30, 1996}
\maketitle

\mediumtext
\begin{abstract}
The capacitance of mesoscopic samples depends on their geometry {\it and}
physical properties, described in terms of characteristic time scales. The
resulting ac admittance shows sample to sample fluctuations. Their
distribution is studied here -through a random-matrix model- for a chaotic
cavity capacitively coupled to a backgate: it is obtained from the
distribution of scattering time delays for the cavity, which is found
analytically for the orthogonal, unitary and symplectic universality
classes, one mode in the lead connecting the cavity to the reservoir and no
direct scattering. The results agree with numerical simulations.
\end{abstract}

}]

\narrowtext

The elementary notion of capacitance of a system of conductors, as a
quantity determined solely by the geometry, has to be revised if the
electric field is not completely screened at the surface of the conductors.
In fact, the penetration distance of the field is of the order of the
Thomas-Fermi screening length, which may be appreciable for a mesoscopic
conductor: the standard description of a capacitor in terms of the {\it
geometric
capacitance} $C_e$ (that relates the charge $Q$ on the plate to the
voltage $%
U$ across the capacitor), gives way, in the mesoscopic domain, to a more
complex entity $C_\mu ,$ the {\it electrochemical capacitance} (that relates
$Q$ to the electrochemical potential of the reservoirs), which depends on
the properties of the conductors \cite{markus93JP}. This fact, in turn, has
important consequences for the ac current induced in the system when the
electrochemical potentials are subject to a nonzero-frequency time variation 
\cite{markus93JP}.

The electrochemical nature of the capacitance has been relevant to a number
of experiments \cite{lambe} and has been
discussed theoretically by
several authors \cite{markus93JP,markus93PL94ZP93PRL,natori}.
Remarkably,
it has been found that the resulting ac admittance can be described in terms
of characteristic time scales related to energy derivatives of scattering
matrix elements.

It is well known that, as a result of quantum interference, the dc
conductance of mesoscopic structures shows strong fluctuations as a function
of the Fermi energy or the magnetic field, as well as from sample to sample.
A statistical analysis of this phenomenon has been done, for diffusive
transport in disordered structures, using microscopic perturbative and
macroscopic random-matrix theories \cite{mesbookbeenakker}, and for
ballistic microstructures --cavities in which impurity scattering can be
neglected so that only scattering from the boundaries is important-- whose
classical dynamics is chaotic, using semiclassical, field theoretic and
random-matrix approaches \cite{haroldpier,rodolfo}.

An extension of the above random-matrix studies to include the ac admittance
of mesoscopic structures is the subject of the present investigation.

In this letter we shall confine our discussion to the geometry shown in Fig.
1. In this system there is, of course,
no dc transport, but there may be an ac current, determined by the
admittance \cite{markus93JP,markus93PL94ZP93PRL} 
\begin{equation}
\label{gI(omega)}g^I(\omega )=\frac{%
g(\omega )}{1+\frac i{\omega C_e}g(\omega )} \equiv -i\omega C_\mu +\cdot
\cdot \cdot \; , 
\end{equation}
written in the Thomas-Fermi approximation and to lowest order in the
frequency $\omega $. Here, $g(\omega )$ , $g^I(\omega )$ 
denote the admittance for the noninteracting
and interacting system, respectively, the former being given, for zero
temperature, by  
\begin{eqnarray}
\label{g(omega)}
g(\omega)&&=-i\omega e^2\left\{ \frac 1{2\pi i}Tr\left[ S^{\dagger }(E) 
\frac{\partial S(E)}{\partial E}\right] \right\}+\cdot \cdot \cdot
\nonumber\\
&&=-i\omega e^2N\tau /\Delta +\cdot \cdot \cdot \; . 
\end{eqnarray}
Here, $S(E)$ is the $N\times N$ scattering matrix for the system formed by
the cavity and the lead, $N$ being the number of propagating modes, or open
channels, in the lead; $\Delta $ is the mean level spacing for the cavity
(the inverse of the level density). Following \cite
{wigner55smithbauerZPbauerJP}, we have introduced the{\it \ dimensionless
time delay} 
\begin{equation}
\label{tau}\tau =\frac \Delta {2\pi N}\frac{\partial \theta }{\partial E}
\; , 
\end{equation}
where $\exp (i\theta )=\det S$ . We then write $g^I(\omega )$ of Eq. (\ref
{gI(omega)}) as 
\begin{equation}
\label{gI(alpha)}g^I(\omega )=-i\omega C_e\alpha \ +\cdot \cdot \cdot \; ,  
\end{equation}
where the dimensionless capacitance $\alpha$ is given by 
\begin{equation}
\label{alpha}\alpha =C_\mu /C_e=\frac \tau {\tau +\eta } 
\end{equation}
and 
\begin{equation}
\label{eta}\eta =\frac{C_e}{N\frac{e^2}\Delta } \; . 
\end{equation}
Notice that, for a macroscopic cavity, $\eta \ll 1$, so that $\alpha \approx
1$ and $g^I(\omega )\approx -i\omega C_e$ .

The {\it one-energy} statistical distribution of the $S$ matrix
for ballistic cavities larger than the Fermi wavelength 
has been modelled successfully through an ``equal-a-priori
probability'' ansatz (known as a ``circular ensemble'') 
\cite{haroldpier,rodolfo}, when the classical dynamics is chaotic and 
direct processes through the
microscructure can be neglected, so that, as a result, the averaged $S$
vanishes, $\overline{S}=0$. 
It is clear, though, that the time delay $\tau $ of Eq. (\ref{tau}) is a
{\it two-energy} function and thus requires more information for its
statistical
study. The distribution of $\tau $, $w(\tau )$, has been studied for a
one-dimensional disordered system within the invariant imbedding formalism
in \cite{narendra}. In another approach, an underlying Hamiltonian described
by a Gaussian ensemble was assumed and the problem analyzed using
supersymmetry techniques: the two-point correlation function for the $S$
matrix elements was derived in \cite{lewenkopf}; phaseshift times for
unitary symmetry, $N$ and $\overline{S}$ arbitrary were studied in 
\cite{fyodorov}. Ref. \cite{seba} finds an approximation to $w(\tau )$.
We concentrate, in what follows, on $w(\tau )$ for arbitrary
symmetry (orthogonal, unitary and symplectic, identified as $\beta =1,2$ and
4, respectively), $N=1$ and $\overline{S}=0$: we show that this case can be
treated using an old conjecture by Wigner \cite{wigner,wigner1}: we
believe that the simplicity of the argument is appealing and gives an
interesting perspective to the problem and a unified point of view for
arbitrary $\beta $. We also remark that, for ballistic cavities, the case of
just one open channel, $N=1$, is very relevant from an experimental point of
view, since cases of small $N$ have been realized in the laboratory \cite
{expdots}. We find below
\begin{equation}
\label{w(tau)}w_\beta (\tau )=\frac{\left( \beta /2\right) ^{\beta /2}}{%
\Gamma \left( \beta /2\right) }\frac{\rm{e}^{-\frac \beta {2\tau }}}{\tau
^{\frac{%
\beta +4}2}} \; ,
\end{equation}
where $0\leq \tau <\infty $. For $\beta =2$, this result agrees with that of
Ref. \cite{fyodorov}. 
The main result of the present paper, i.e. the $\beta$ dependent
distribution of
the dimensionless capacitance  
$\alpha$ [$\alpha$ is related to the ac admittance via Eq.
(\ref{gI(alpha)})],
then follows as
\begin{equation}
\label{p(alpha)}p_{\beta ,\eta }(\alpha )=\frac{\left( \beta /2\right)
^{\beta /2}}{\Gamma \left( \beta /2\right) }\frac{\left( 1-\alpha \right)
^{\beta /2}}{\eta ^{\frac{\beta +2}2}\alpha ^{^{\frac{\beta
+4}2}}}\rm{e}^{-\beta 
\frac{1-\alpha }{2\eta \alpha }} \; ,
\end{equation}
for 0$\leq \alpha \leq 1$. A plot of $p_{1,\eta }(\alpha )$ for various
values of $\eta $ is presented in Fig. 2. For a macroscopic cavity, $\eta
\rightarrow 0$ and $p_{\beta ,\eta }(\alpha )\rightarrow \delta (1-\alpha )$.
We now derive the distribution of time delays, Eq. (\ref{w(tau)}).

We write $S$ for $N=1$ as 
\begin{equation}
\label{SKtheta}S(E)=\frac{1+iK(E)}{1-iK(E)}=\rm{e}^{i\theta (E)} \; . 
\end{equation}
For pure resonance scattering the $K$ function can be given the
sum-over-resonance form \cite{wigner,wigner1} 
\begin{equation}
\label{K(E)}K(E)=\sum_\lambda \frac{\Gamma _\lambda }{E_\lambda -E} \; ,
\end{equation}
where the ``widths'' $\Gamma _\lambda $ for a given symmetry class $\beta $
can be written in terms of real amplitudes $\gamma _\lambda ^{(i)}$ as 
\begin{equation}
\label{Gamma}\Gamma _\lambda =\sum_{i=1}^\beta \left[ \gamma _\lambda
^{(i)}\right] ^2 \; . 
\end{equation}

The quantity $\theta ^{^{\prime }}(E)/2=h(E)$ was studied extensively by
Wigner \cite{wigner,wigner1}; it is called the ``invariant derivative'',
because it remains invariant under the transformation 
\begin{equation}
\label{Kphi}K_\phi =\frac{K+\tan \phi }{1-K\tan \phi } \; , 
\end{equation}
$\phi $ being a constant; since $K=\tan (\theta /2)$, (\ref{Kphi}) takes 
$\theta /2$ to $\theta /2+\phi$, and hence $S$ to $e^{i\phi}Se^{i\phi}$.
Both $K$ and its transforms have the form $K=\tan \int_c^Eh(E)dE$, 
$c$ being different for different transforms. 
Starting from one pole $E_1$ of $K$ , one can obtain the next one by
determining the abscissa $E_2$ so that the area under $h(E)$ between $E_1$
and $E_2$ is $\pi .$ Moreover, at a pole $E_\lambda $ we have $\Gamma
_\lambda =1/h(E_\lambda )$. These relations are shown in Fig. 3. The levels
and widths of the transforms of $K$ can be obtained by a similar
construction, starting at another abscissa.

From (\ref{SKtheta},\ref{K(E)}) we find the energy average of $S(E)$ as 
\begin{equation}
\label{Sbar}\overline{S(E)}=S(E+iI)=\frac{1-t}{1+t} \; ,
\end{equation}
where $I\rightarrow \infty $ \cite{feshbach} and $t=\pi \overline{\Gamma }%
/\Delta $. For $\overline{S(E)}=0$ [circular ensemble, invariant under 
(\ref{Kphi})], we have $t=1$. In this case (referred to
in Ref. \cite{wigner,wigner1}
as that of a ``normalized'' $R$ function) Wigner proposes the

{\it Conjecture}: the statistical distributions of level spacings and
residues are invariant under the transformation (\ref{Kphi}).

The above statemenent is a ``conjecture", not a ``theorem", and it is not 
clear, a priori,
for what distributions, if any, it is fulfilled. Wigner, in his papers,
proposes it for ``most statistical distributions". 
The conjecture, in relation with the residue distribution, was verified
numerically for the case in which the energy levels entering Eq. (\ref{K(E)}%
) are constructed from a  Gaussian Orthogonal, Unitary or
Symplectic Ensemble, and the $\gamma _\lambda ^{(i)}$ of Eq. (\ref{Gamma}) 
as independent Gaussian variables: the residue distribution was
found to remain invariant, within the statistical error bars of the
numerical simulation. On the other hand, the
conjecture is seen, in our numerical studies, to be violated for a
spectrum of statistically independent energy levels following a
Poisson distribution.

Call $Q\left( h\right) $ the probability density of the inverse widths $%
h(E_\lambda )=h_\lambda $ and $P(h)$ the probability density of $h$ across
the energy axis, irrespective of whether we are at resonance or not; $P(h)$ 
is related to $w(\tau)$ as $w(\tau)=(\pi /\Delta)P(\pi \tau /\Delta)$.
Assuming the above conjecture, Ref. \cite{wigner1} shows that 
\begin{equation}
\label{PvsQ}P(h)=\frac \pi {h\Delta }Q(h) \; . 
\end{equation}
This relation can be understood by means of a very simple argument.
Consider, for one given $K$, the following {\it level-average } 
\begin{equation}
\label{avf(h)}\left\langle f(h)\right\rangle _\lambda =\frac 1m\sum_{\lambda
=1}^mf(h_\lambda ) \; ,
\end{equation}
for an arbitrary function $f$. From Fig. 3 we see that we cannot replace the
sum in this equation by an integral. However, using the transformation (\ref
{Kphi}) we can construct ``replicas'' of $K$, all having the same
distribution of $h_\lambda $; we do this $n$ times, in such a way that the
area between two successive levels is subdivided into $n$ strips of area $%
\pi /n$ each. Now we have a fine mesh, the sum over which can be
approximated by an integral, using a density $nh/\pi $, since the base of
one of the above strips, at the place where $h$ is the local value of the
curve, is $\pi /nh$. We then arrive at the above relation (\ref{PvsQ}).

If we use, in (\ref{PvsQ}), the variable $u=\pi /h\Delta$ (and denote the 
distributions with a hat), we have 
\begin{equation}
\label{P(u)}\widehat{P}(u)=u\widehat{Q}(u) \; . 
\end{equation}
On the LHS, $u$ can be thought of in terms of $\tau $ of Eq. (\ref{tau})
as $%
u=1/\tau $; at resonance, $u$ takes the value $u_\lambda =\pi \Gamma
_\lambda /\Delta $, which is the relevant variable on the RHS of Eq. (\ref
{P(u)}). Thus, knowing the distribution of widths $\widehat{Q}(u)$, Eq.
(\ref{P(u)})
allows finding $\widehat{P}(u)$ \cite{note}.

For the three universality classes $\beta =1,2,4$ and independent Gaussian
variables 
$\gamma _\lambda ^{(i)}$, the distribution $\widehat{Q}(u)$ is the
chi-square 
distribution function with $\beta$ degrees of freedom,

\begin{equation}
\label{Q(u)}\widehat{Q}_\beta (u)=\frac{\left( \beta /2\right) ^{\beta
/2}}{\Gamma
\left( \beta /2\right) }u^{\frac{\beta -2}2}\rm{e}^{-\frac \beta 2u} \; ;
\end{equation}
Eq. (\ref{P(u)}) then gives $\widehat{P}(u)$, from which we find the
distribution of
time delays $w_\beta (\tau )$ of Eq. (\ref{w(tau)}). We notice the
remarkable fact that, while $w_\beta (\tau )$ certainly depends on the
distribution of widths, {\it other characteristics of the spectrum become
lumped together in the invariance property contemplated in Wigner's
conjecture}. 

A numerical verification (using the simulation explained above in 
relation with Wigner's conjecture) 
of $w_\beta (\tau )$ of Eq. (\ref{w(tau)}) is shown
in Fig. 4 for the three symmetry classes $\beta =1,2,4$: in all cases the
agreement is
seen to be very good.

To summarize, we have found the statistical distribution of capacitances $%
p_{\beta ,\eta }(\alpha )$, Eq. (\ref{p(alpha)}), $\alpha $ being defined in
Eqs. (\ref{gI(alpha)},\ref{alpha}), for the system shown in Fig. 1, whose
essential element is a mesoscopic capacitor. The plate coupled to the
backgate is a chaotic cavity; the experimentally relevant situation of one
open channel ($N=1$) is considered and the possibility of direct reflection
by the cavity is neglected. The essential ingredient that is needed is the
statistical distribution $w_\beta \left( \tau \right) $, Eq. (\ref{w(tau)}),
of time delays $\tau $ associated with the scattering from the cavity. It is
shown that $w_\beta \left( \tau \right) $ can be obtained in a very simple
way from a conjecture by Wigner, whose validity, in turn, is verified
numerically for the three symmetry classes: orthogonal, unitary and 
symplectic. The resulting $w_\beta \left( \tau
\right) $ compares very well with the results of numerical simulations, for
the three classes. The statistical analysis of the admittance of mesoscopic 
conductors provides additional information on such systems not contained in
the investigation of dc transport properties, and thus points to an
interesting 
avenue of future research.

Part of this work was supported by the DGAPA, M\'exico, and the Swiss
National 
Science Foundation.

\begin{figure}
\caption{Mesoscopic capacitor: A cavity (thick line)
is connected via a perfect lead to reservoir 1 and
capacitively coupled to a macroscopic backgate (thin line)
connected to reservoir 2. The 
cavity is ballistic and its classical dynamics is chaotic. 
}
\end{figure}

\begin{figure}
\caption{The probability density of $\alpha$ -the ratio of the
electrochemical to the geometric capacitance- for the orthogonal case
(Eq. (8)), for a number of values of $\eta$.
}
\end{figure}

\begin{figure}
\caption{The invariant derivative $h(E)$. The  $E_\lambda$ are the poles of
$K(E)$ and the
$\Gamma _\lambda$ the corresponding widths. Replicas of $K$ with the
same width distribution (according to Wigner's conjecture) are generated 
via the transformation (12) and used to subdivide the area between 
successive levels into $n$ strips.
}
\end{figure}

\begin{figure}
\caption{The distribution $W_{\beta}(\tau)$ of time delays 
for one channel  
and in the absence of direct processes, for the (a) orthogonal, (b) unitary 
and (c) symplectic universality classes. The dotted curves are proportional 
to the theoretical probability density given by Eq. (7). The points 
with the finite-sample error bar are the results of the numerical simulation 
described in the text: $200$-dimensional matrices were used in the three 
cases. The agreement is excellent.
}
\end{figure}


\begin{thebibliography}{99}
\bibitem{note}  The integral of the LHS\ of ($\ref{P(u)})$ is 1 from
normalization; the fact that the integral of the RHS is also 1 means $%
\left\langle u\right\rangle _{\widehat{Q}} =1$, which is consistent with
$t=1$, as required
right after Eq. (\ref{Sbar}). Similarly, from the normalization of 
$\widehat{Q}(u)$, Eq. (\ref{P(u)}) shows that $\left\langle 1/u\right\rangle
_{\widehat{P}} =1$, and hence $\left\langle \tau \right\rangle _w = 1$.
\end{thebibliography}
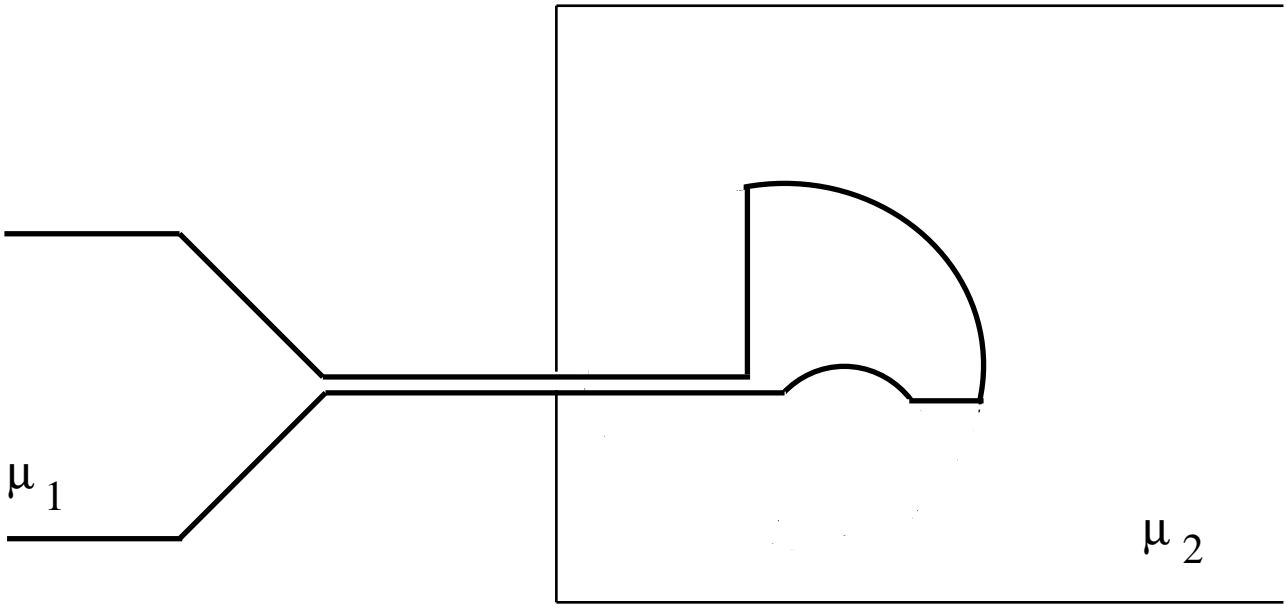
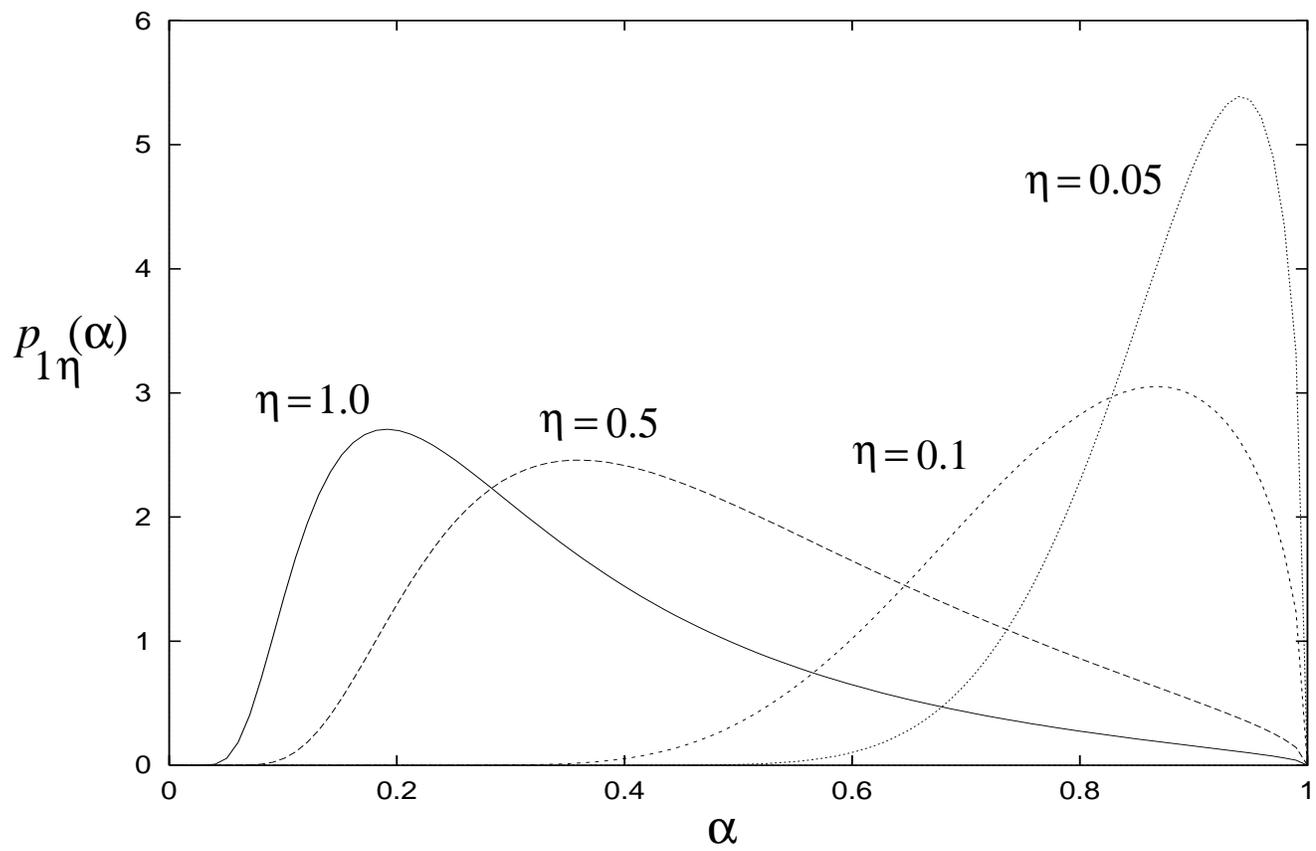
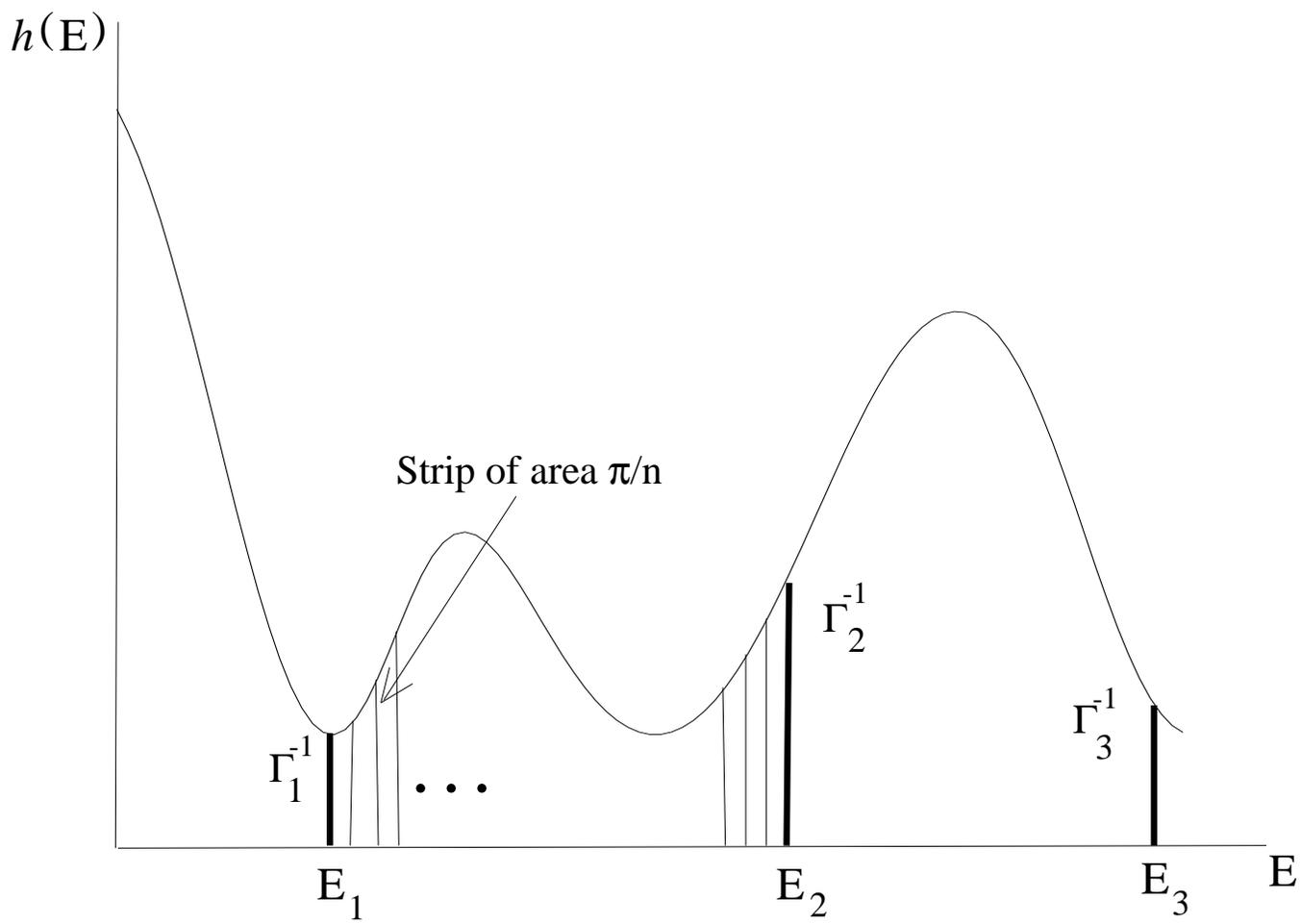
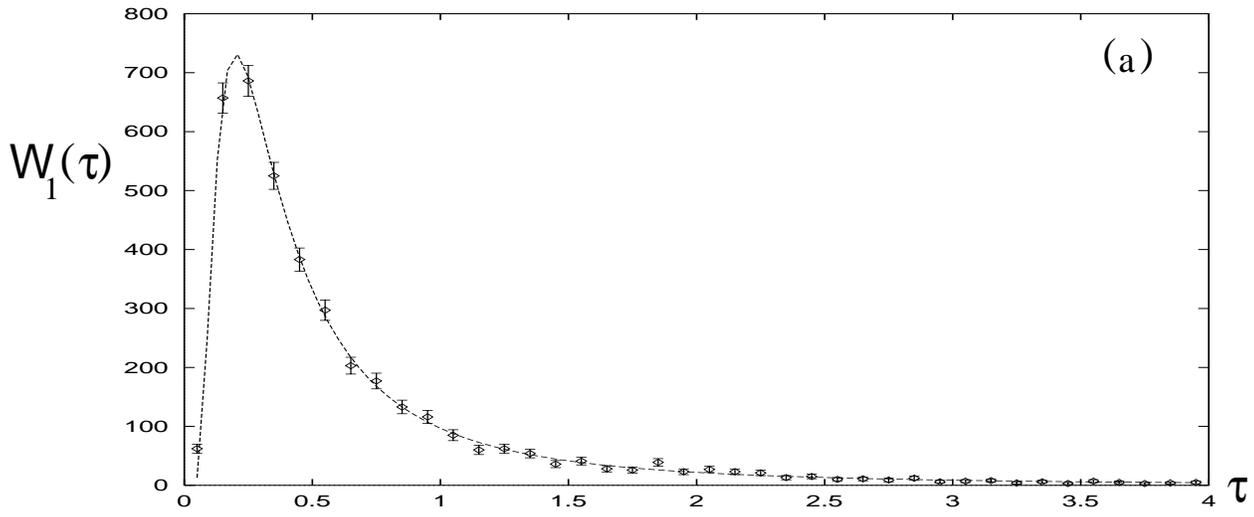
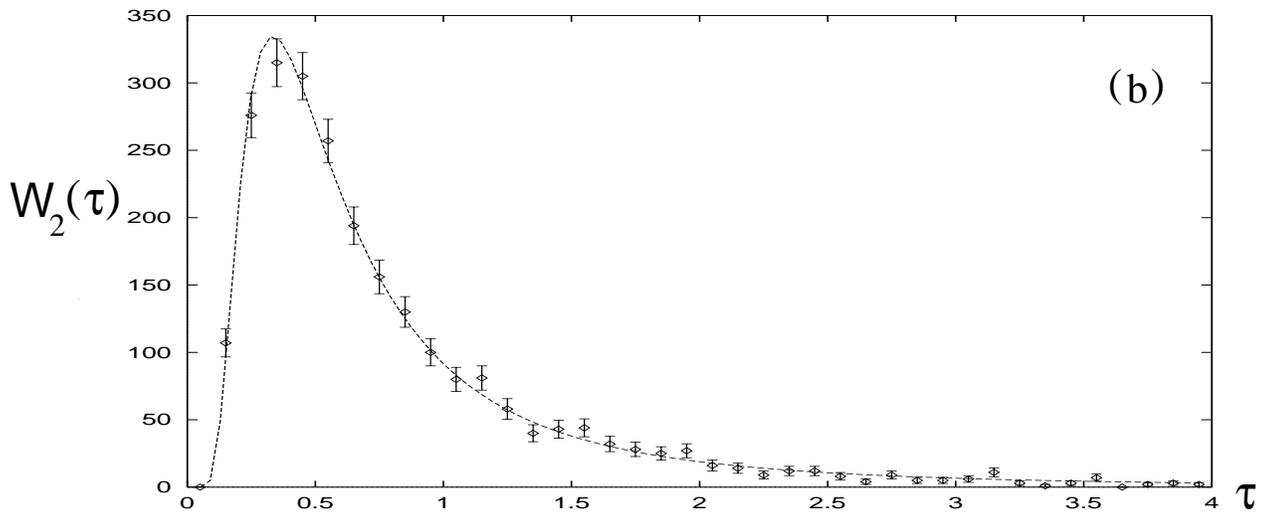
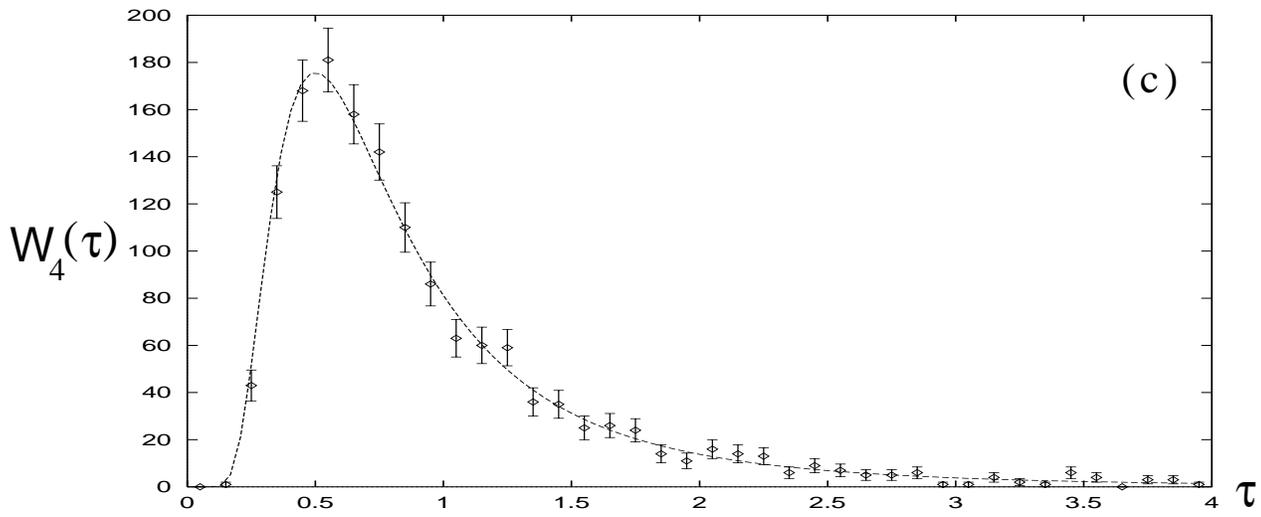
\end{document}